\documentclass{entcs} \usepackage{entcsmacro}
\usepackage{pst-coil}
\usepackage{epsfig}
\usepackage{url}
\usepackage[normalem]{ulem}
\usepackage{graphpap}
\usepackage{amssymb}
\usepackage{color,pstcol}
\usepackage{pstricks}
\usepackage{pst-text}
\usepackage{pst-node}
\usepackage{pst-rel-points}
\usepackage{mathptm}
\usepackage{longtable}

\protect{\newlength{\TAL}}
\protect{\newcounter{programline}}
\newcommand{\DEFINE}{\mbox{\sf\em define}}
\newcommand{\CONS}{\mbox{\sf\em cons}}
\newcommand{\CAR}{\mbox{\sf\em car}}
\newcommand{\CDR}{\mbox{\sf\em cdr}}
\newcommand{\IF}{\mbox{\sf\em if}}
\newcommand{\LET}{\mbox{\sf\em let}}
\newcommand{\NIL}{\mbox{\sf\em Nil}}

\newcommand{\UNL}[1]{\\ & \hspace*{#1\TAL}
}
\newcommand{\UFL}{ & }
\newenvironment{uprogram}{\begin{tabular}{@{}r@{}l@{}}}{\\\end{tabular}}

\newcommand{\myrule}{}

\newcommand{\scmin} {\mbox{\sf\em in}}
\newcommand{\scmnull}{\mbox{\sf\em null?}}
\newcommand{\scmpair}{\mbox{\sf\em pair?}}
\newcommand{\scmprim}{\ensuremath{\mathsf{+}}}

\newcommand{\acar}{\ensuremath{\mathbf{0}}}
\newcommand{\acdr}{\ensuremath{\mathbf{1}}}
\newcommand{\bcar}{\ensuremath{\bar\acar}}
\newcommand{\bcdr}{\ensuremath{\bar\acdr}}
\newcommand{\acarset}{\ensuremath{\lbrace\acar\rbrace}}
\newcommand{\acdrset}{\ensuremath{\lbrace\acdr\rbrace}}
\newcommand{\bcarset}{\ensuremath{\lbrace\bcar\rbrace}}
\newcommand{\bcdrset}{\ensuremath{\lbrace\bcdr\rbrace}}
\newcommand{\epsilonset}{\ensuremath{\lbrace\epsilon\rbrace}}

\newcommand{\xf}[2]{\ensuremath{\xfonly_{\!\!#1}^{\,#2}}}
\newcommand{\xfonly}{\ensuremath{\mathcal{XF}}}
\newcommand{\find}[2]{\ensuremath{\mathcal{I}_{#1}^{#2}}}
\newcommand{\fdep}[2]{\ensuremath{\mathcal{D}_{#1}^{#2}}}

\newcommand{\xp}[2]{\ensuremath{\xponly_{\!\!#1}^{\,#2}}}
\newcommand{\xponly}{\ensuremath{\mathcal{XP}}}

\newcommand{\xe}[3]{\ensuremath{\xeonly(#1, #2, #3)}}
\newcommand{\xeonly}{\ensuremath{\mathcal{XE}}}

\newcommand{\lv}{\ensuremath{\mathcal{L}}}
\newcommand{\ptexp}[2]{#2}

\newcommand{\plus}{\cup}

\newcommand{\rightk}[1]{\ensuremath{\stackrel{\scriptstyle #1}{\rightarrow}}}
\newcommand{\rightstar}{\rightk{\star}}
\newcommand{\eqdef}{\ensuremath{=}}

\newcommand{\append}{\mbox{\sf app}}

\newcommand{\fnappend}{\ensuremath{\scriptstyle \append}}
\newcommand{\lista}{\mbox{\sf list1}}
\newcommand{\listb}{\mbox{\sf list2}}

\newcommand{\nfa}{\ensuremath{\mathbf{N}}}
\newcommand{\nfabar}{\ensuremath{\overline\nfa}}

\newcommand{\prim}{{P}}
\newcommand{\exit}{{\sf pgm}}
\newcommand{\gram}{\ensuremath{G}}

\newcommand{\TwoCells}[2]{%
\psset{unit=.25mm}
\begin{pspicture}(0,-2)(36,18)
\psframe(0,-5)(36,15)
\psline(18,-4)(18,15)
\putnode{z}{origin}{9}{5}{\rnode{#1}{}}
\putnode{z}{origin}{27}{5}{\rnode{#2}{}}
\end{pspicture}%
}

\def\lastname{Karkare, Khedker and Sanyal}

\begin{document}
\begin{frontmatter}
  \title{Liveness of Heap Data for Functional Programs}
  \author{Amey Karkare\thanksref{Infosys}}
  \author{Uday Khedker}
  \author{Amitabha Sanyal}
  \address{{\tt \{karkare,uday,as\}@cse.iitb.ac.in}\\
	Department of CSE, IIT Bombay\\ Mumbai, India}
  \thanks[Infosys]{Supported by Infosys Technologies Limited,
    Bangalore, under Infosys Fellowship Award.}

\begin{abstract} 
Functional  programming  languages  use  garbage collection  for  heap
memory  management.  Ideally,  garbage collectors  should  reclaim all
objects  that are {\em  dead} at  the time  of garbage  collection. An
object  is  dead  at  an  execution  instant if  it  is  not  used  in
future. Garbage  collectors collect only  those dead objects  that are
not reachable from any program  variable. This is because they are not
able  to  distinguish between  reachable  objects  that  are dead  and
reachable objects that are live.

In this  paper, we  describe a static  analysis to  discover reachable
dead  objects in  programs  written in  first-order, eager  functional
programming languages.  The results of  this technique can be  used to
make  reachable  dead objects  unreachable,  thereby allowing  garbage
collectors to reclaim more dead objects.
\end{abstract}

\begin{keyword}
  Compilers, Liveness, Garbage Collection, Memory Management, Data
  Flow Analysis, Context Free Grammars
\end{keyword}

\end{frontmatter}

\section{Introduction}
\label{sec:intro_live}
Garbage  collection  is an  attractive  alternative  to manual  memory
management because it frees  the programmer from the responsibility of
keeping  track of  object  lifetimes. This  makes  programs easier  to
implement,  understand  and  maintain.   Ideally,  garbage  collectors
should reclaim all objects that are  {\em dead} at the time of garbage
collection. An  object is dead  at an execution  instant if it  is not
used in future.  Since garbage  collectors are not able to distinguish
between reachable objects that are live and reachable objects that are
dead, they conservatively approximate the liveness of an object by its
reachability  from a  set of  locations called  {\em root  set} (stack
locations        and        registers        containing        program
variables)~\cite{jones96gc}.  As a  consequence, many dead objects are
left uncollected.   This has been  confirmed by empirical  studies for
Haskell~\cite{rojemo96lag},  Scheme~\cite{karkare06effectiveness}  and
Java~\cite{shaham00gc,shaham01heap,shaham02estimating}.

Compile  time analysis  can help  in distinguishing  reachable objects
that are  live from reachable objects  that are dead. This  is done by
detecting unused  references to  objects.  If an  object is dead  at a
program point, none  of its references are used  by the program beyond
that program point.  If every unused reference is  nullified, then the
dead  objects may  become unreachable  and may  be claimed  by garbage
collector.

\begin{figure}
  \begin{center}
    \begin{tabular}{cc}
      {\sf
	\renewcommand{\arraystretch}{1}{
	  \begin{uprogram}
	  \UFL\ \hspace*{-1\TAL} (\DEFINE\ (\append\  \lista\ \listb)
	  \UNL{0}  $\pi_1\!:$(\IF\ $\pi_2\!:$(\scmnull\ $\pi_3\!:$\lista)
	  \UNL{1}      $\pi_4\!:$\listb
	  \UNL{1}      $\pi_5\!:$(\CONS\ $\pi_6\!:$(\CAR\  $\pi_7\!:$\lista)
	  \UNL{3}          \ \ \ $\pi_8\!:$(\append\ $\pi_9\!:$(\CDR\  $\pi_{10}\!:$\lista)
	  \UNL{6}          \                        $\pi_{11}\!:$\listb))))
	  \UNL{0}(\LET\ z  $\leftarrow$(\CONS\ (\CONS\ $4$ (\CONS\ $5$ \NIL))
	  \UNL{3}\ \ \  (\CONS\ $6$ \NIL)) \scmin
	  \UNL{0} (\LET\ y  $\leftarrow$ (\CONS\ $3$ \NIL) \scmin
	  \UNL{1}\  $\pi_{12}:$(\LET\ w $\leftarrow$ $\pi_{13}:$(\append\ y z)\ \scmin
	  \UNL{4}               $\pi_{14}:$(\CAR\ (\CAR\ (\CDR\ w))))))
	\end{uprogram}
      }} 
      &
      \raisebox{-25mm}{\scalebox{.9}{
      \psset{unit=1mm}
      \psset{linewidth=.3mm}
      \begin{pspicture}(0,0)(73,60)
	\putnode{o}{origin}{13}{50}{\TwoCells{o1}{o2}}
	\putnode{a}{o}{-10}{-15}{\psframebox{3}}
	\putnode{b}{o}{10}{-15}{\psframebox[linestyle=none,framesep=.5]{\NIL}}
	\ncline[offsetB=-.5,nodesepB=.1]{*->}{o1}{a}
	\ncline[offsetB=-.5,nodesepB=.1]{*->}{o2}{b}
	\putnode{y}{o}{-14}{8}{\psframebox[linestyle=none,framesep=.5]{y}}
	\nccurve[nodesepB=-.2,angleA=330,angleB=120]{->}{y}{o}
	\aput[-3.5](.5){\scalebox{1.2}{\psframebox[framesep=.2,linestyle=none,fillstyle=solid,
		  fillcolor=white]{$\times$}}}
	\putnode{c}{o}{25}{0}{\TwoCells{c1}{c2}}
	\putnode{d}{c}{10}{-10}{\TwoCells{d1}{d2}}
	\putnode{e}{d}{-13}{-12}{\TwoCells{e1}{e2}}
	\putnode{f}{d}{13}{-12}{\TwoCells{f1}{f2}}
	\ncline[nodesepB=-.5]{*->}{c2}{d}
	\ncline[nodesepB=-.5,linewidth=.7]{->}{c2}{d}
	\nccurve[ncurv=1,angleA=270,angleB=330]{*->}{c1}{a}
	\aput[-3.5](.2){\scalebox{1.2}{\psframebox[framesep=.2,linestyle=none,fillstyle=solid,
	      fillcolor=white]{$\times$}}}
	\nccurve[nodesepB=-.5,angleA=240,angleB=70]{*->}{d1}{e}
	\nccurve[nodesepB=-.5,angleA=240,angleB=70,linewidth=.7]{->}{d1}{e}
	\nccurve[nodesepB=-.5,angleA=300,angleB=110]{*->}{d2}{f}
	\aput[-3.5](.5){\scalebox{1.2}{\psframebox[framesep=.2,linestyle=none,fillstyle=solid,
	      fillcolor=white]{$\times$}}}
	\putnode{w}{c}{-8}{8}{\psframebox[linestyle=none,framesep=.2]{w}}
	\putnode{ww}{c}{15}{8}{\psframebox[linestyle=none,framesep=.2]{z}}
	\nccurve[nodesepB=-.2,angleA=330,angleB=120,linewidth=.7]{->}{w}{c}
	\putnode{g}{e}{-8}{-12}{\psframebox{4}}
	\putnode{h}{e}{8}{-14}{\TwoCells{h1}{h2}}
	\putnode{i}{f}{-8}{-11}{\psframebox{6}}
	\putnode{j}{f}{8}{-11}{\psframebox[linestyle=none,framesep=.5]{\NIL}}
	\ncline[offsetB=-.5,nodesepB=.1]{*->}{e1}{g}
	\ncline[offsetB=-.5,nodesepB=.1,linewidth=.7]{->}{e1}{g}
	\ncline[offsetB=-.5,nodesepB=-.3]{*->}{e2}{h}
	\aput[-3.2](.5){\scalebox{1.2}{\psframebox[framesep=.1,linestyle=none,fillstyle=solid,
	      fillcolor=white]{$\times$}}}
	\ncline[offsetB=-.5,nodesepB=.1]{*->}{f1}{i}
	\ncline[offsetB=-.5,nodesepB=.1]{*->}{f2}{j}
	\nccurve[nodesepB=-.2,angleA=270,angleB=90]{->}{ww}{d}
	\aput[-3.2](.4){\scalebox{1.2}{\psframebox[framesep=.1,linestyle=none,fillstyle=solid,
	      fillcolor=white]{$\times$}}}
	\putnode{k}{h}{-8}{-11}{\psframebox{5}}
	\putnode{l}{h}{8}{-11}{\psframebox[linestyle=none,framesep=.5]{\NIL}}
	\ncline[offsetB=-.5,nodesepB=.1]{*->}{h1}{k}
	\ncline[offsetB=-.5,nodesepB=.1]{*->}{h2}{l}
      \end{pspicture}}} \\
      (a) Example program.&
      \renewcommand{\arraystretch}{.9}{\begin{tabular}[t]{ll}
	(b) Memory graph at $\pi_{14}$.\\
	{\white (b) }Thick edges denote live links. \\
	{\white (b) }Edges marked $\times$ can be nullified.
      \end{tabular}}
    \end{tabular}
  \end{center}
\caption{Example Program and its Memory Graph.}\label{fig:mot-exmp}    
\myrule
\end{figure}
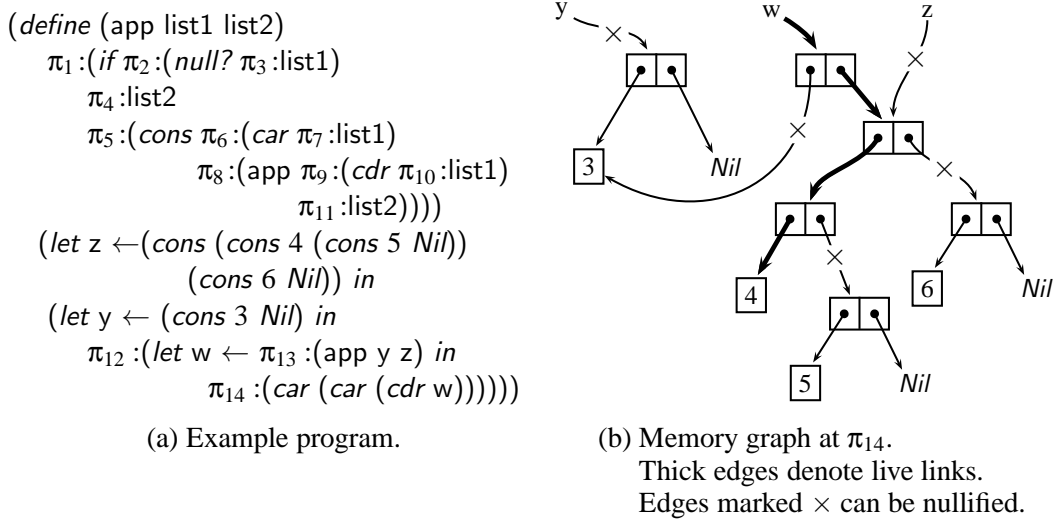

\newcommand{\ww}{{\sf w}}
\newcommand{\xx}{{\sf x}}
\newcommand{\yy}{{\sf y}}
\newcommand{\zz}{{\sf z}}

\begin{example}\label{exmp:mot-exmp}
Figure~\ref{fig:mot-exmp}(a)  shows an  example  program.  The  label
$\pi$ of an expression $e$  denotes the program point just before the
evaluation of $e$.
At a given program point, the heap memory can be viewed as a (possibly
unconnected) directed  acyclic graph  called {\em memory  graph}.  The
locations in the  root set form the entry nodes  for the memory graph.
Figure~\ref{fig:mot-exmp}(b)  shows the  memory  graph at  $\pi_{14}$.
Each \CONS\ cell is an intermediate {\em node} in the graph.  Elements
of basic data types and the 0-ary constructor \NIL\ form leaf nodes of
the graph.   They are  assumed to be  boxed, i.e.\ stored  in separate
heap cells and are accessed through references. The edges in the graph
are called {\em links}.

If we consider the execution  of the program starting from $\pi_{14}$,
the links  in the memory graph  that are traversed are  shown by thick
arrows.  These links are {\em  live} at $\pi_{14}$. Links that are not
live  can   be  nullified  by  the  compiler   by  inserting  suitable
statements. If an object  becomes unreachable due to nullification, it
can be collected by the garbage collector.

In  the figure,  the links  that  can be  nullified are  shown with  a
$\times$.  Note that  a link need not be  nullified if nullifying some
other link makes it unreachable from  the root set.  If a node becomes
unreachable  from the  root set  as  a consequence  of nullifying  the
links,  it will  be collected  during the  next invocation  of garbage
collector.
\qed
\end{example}

In this example,  starting at $\pi_{14}$, there is  only one execution
path.  In general, there  could be  multiple execution  paths starting
from a  program point  $\pi$. The liveness  information at $\pi$  is a
combination  of  liveness   information  along  every  execution  path
starting at $\pi$.

In this paper,  we describe a static analysis  for programs written in
first-order,  eager functional  programming  languages.  The  analysis
discovers live references at every program point, i.e.\ the references
that may  be used  beyond the  program point in  any execution  of the
program.  We use context free grammars as a bounded representation for
the set of live references.  The result of the analysis can be used by
the compiler to decide whether a given reference can be nullified at a
given program point.  Our analysis is context-sensitive yet modular in
that a function is analyzed only once.

The rest of the paper is organized as follows:
Section~\ref{sec:lang}  describes  the language  used  to explain  our
analysis along with the basic concepts and notations.
The  analysis   in  Section~\ref{sec:method}  captures   the  liveness
information of  a program as a  set of equations. The  method to solve
these equations is given in Section~\ref{sec:solve-eqns}.
Section~\ref{sec:apps} describes how the result of the analysis can be
used to nullify unused references.
Finally,   we   compare   our    approach   with   related   work   in
Section~\ref{sec:rel-work} and conclude in Section~\ref{sec:concl}.

\section{Language, Concepts and Notations}
\label{sec:lang}

\begin{figure}[t]
  \begin{center}
  \renewcommand{\arraystretch}{1}{
    \begin{eqnarray*}
      p & ::= & d_1 \ldots d_n \; e_1 \;\;\;\;\;\;\;\;\;
      \;\;\;\;\;\;\;\;\;\;\;\;\;\;
      \mbox{\em --- program}\\
      d & ::= & (\DEFINE\; (f v_1 \; \ldots \;v_n)\; e_1) \;\;\;\;\;\; 
      \mbox{\em --- function definition} \\ 
      e & ::= & \;\;\;\;\;\;\;\;\;\;\;\;\;\;\;\;\;\;\;\;\;\;\;\;\;\;
      \;\;\;\;\;\;\;\;\;\;\;\;\;\;
      \mbox{\em --- expression}\\
      & & \begin{array}{ll}
        \;\; \kappa & \mbox{\em --- constant }\\
        \mid v & \mbox{\em --- variable} \\
	\mid \NIL & \mbox{\em --- primitive constructor} \\
	\mid (\CONS\; e_1\; e_2) & \mbox{\em --- primitive constructor} \\
	\mid (\CAR\; e_1) & \mbox{\em --- primitive selector} \\
	\mid (\CDR\; e_1) & \mbox{\em --- primitive selector} \\
	\mid (\scmpair\; e_1) & \mbox{\em --- primitive tester} \\
	\mid (\scmnull\; e_1) & \mbox{\em --- primitive tester} \\
	\mid (\scmprim\; e_1\; e_2) & \mbox{\em --- generic primitive} \\
	\mid (\IF\; e_1\; e_2\; e_3) & \mbox{\em --- conditional} \\
	\mid (\LET\; v_1 \leftarrow e_2\; \scmin\; e_3)
	& \mbox{\em --- let binding} \\
	\mid (f\; e_1\;\ldots\; e_n) & \mbox{\em --- function application}
      \end{array}
    \end{eqnarray*}}
  \end{center}
  \caption{The syntax of our language}\label{fig:lang}
  \myrule
\end{figure}

The syntax of our language is described in Figure~\ref{fig:lang}.  The
language  has call-by-value  semantics. The  argument  expressions are
evaluated from left to right.  We assume that variables in the program
are renamed  so that  the same  name is not  defined in  two different
scopes.

For notational convenience, the  left link (corresponding to the \CAR)
of  a   \CONS\  cell  is  denoted   by  \acar\  and   the  right  link
(corresponding to the \CDR) is  denoted by \acdr.  We use $e.\acar$ to
denote the  link corresponding to  $(\CAR\; e)$ for an  expression $e$
(assuming $e$  evaluates to a list)  and $e.\acdr$ to  denote the link
corresponding to  $(\CDR\; e)$.  A composition of  several {\CAR}s and
{\CDR}s     is    represented    by     a    string     $\alpha    \in
\lbrace\acar,\acdr\rbrace^*$.   If an  expression $e$  evaluates  to a
\CONS\  cell then  $e.\epsilon$ corresponds  to the  reference  to the
\CONS\ cell.

\newcommand{\loc}[1]{\ensuremath{[#1]}}

%
For an expression  $e$, let $\loc{e}$ denote the  location in the root
set holding the value of $e$.
Given a  memory graph, the string  $e.\alpha$ describes a  path in the
memory graph that starts at $\loc{e}$.  We call the string $e.\alpha$
an  {\em access  expression\/},  the string  $\alpha$  an {\em  access
pattern\/}, and  the path  traced in the  memory graph an  {\em access
path}.    In    Figure~\ref{fig:mot-exmp},   the   access   expression
\ww.\acdr\acar\acar\ represents the access  path from \ww\ to the node
containing the value $4$.  Most often, the memory graph being referred
to  is clear  from  the context,  and  therefore we  shall use  access
expressions to refer  to access paths.  When we use  an access path to
refer to a link  in the memory graph, it denotes the  last link in the
access path.  Thus, \ww.\acdr\acar\acar\  denotes the link incident on
the node containing the value $4$.
If $\sigma$ denotes  a set of access patterns,  then $e.\sigma$ is the
set  of  access  paths   rooted  at  $\loc{e}$  and  corresponding  to
$\sigma$. i.e.
\renewcommand{\arraystretch}{.99}
\begin{eqnarray*}
  e.\sigma & \eqdef & \lbrace e.\alpha \mid \alpha \in \sigma \rbrace 
\end{eqnarray*}

A link  in a memory graph  is live at a  program point $\pi$  if it is
used in  some path from $\pi$ to  the program exit. An  access path is
defined to be live if its last link is live.
In  Example~\ref{fig:mot-exmp},  the  set  of  live  access  paths  at
$\pi_{14}$   is  $\lbrace  \ww.\epsilon,   \ww.\acdr,  \ww.\acdr\acar,
\ww.\acdr\acar\acar, \zz.\acar, \zz.\acar\acar\rbrace$.  Note that the
access paths \zz.\acar\ and \zz.\acar\acar\ are live at $\pi_{14}$ due
to  sharing.  We do  not discover  the liveness  of such  access paths
directly. Instead, we assume that an optimizer using our analysis will
use alias analysis to discover liveness due to sharing.

The end result  of our analysis is the  annotation of every expression
in the program with a set of access paths rooted at program variables.
We  call  this  {\em  liveness  environment\/},  denoted~\lv.   This
information  can  be  used  to  insert  nullifying  statements  before
expressions.

The  symbols  \acar\  and  \acdr\  extend the  access  patterns  of  a
structure to  describe the access  patterns of a larger  structure. In
some situations, we  need to create access patterns  of a substructure
from the access  patterns of a larger structure.  For this purpose, we
extend our alphabet  of access patterns to include  symbols \bcar\ and
\bcdr.  The following example motivates the need for these symbols.

\begin{example}\label{exmp:motiv-bar}
  Consider the expression at program point $\pi_1$ in
  \begin{center}
    {\sf $\pi_1\!:${(\LET\ w $\leftarrow$
	                    $\pi_2\!:${(\CONS\ x y)} \scmin\ 
			    $\pi_3\!:${$\cdots$})}}
  \end{center}
  Assuming  \mbox{$\lv_{\pi_3} = \lbrace\ww.\alpha\rbrace$},  we would
  like to find out which reference  of the list \xx\ and \yy\ are live
  at $\pi_1$.   Let \xx.$\alpha'$ be  live at $\pi_1$.  Then,  the two
  possible cases are:
  \begin{itemize}
  \item If  $\alpha = \acdr\beta$ or  $\alpha = \epsilon$,  no link in
    the structure rooted at \xx\ is  used. We use $\bot$ to denote the
    access  pattern describing  such  a situation.   Thus, $\alpha'  =
    \bot$.
  \item  If  $\alpha  =  \acar\beta$  then  the  link  represented  by
    $\ww.\alpha$  that is  \xx\  rooted  and live  at  $\pi_1$ can  be
    represented by $\xx.\beta$. Thus, $\alpha' = \beta$.
  \end{itemize}
  This  relation  between  $\alpha$  and  $\alpha'$  is  expressed  by
  $\alpha'  =  \bcar\alpha$.  \bcdr\  can be  interpreted
  similarly. \qed
\end{example}

With the  inclusion of  \bcar, \bcdr\ and  $\bot$ in the  alphabet for
access patterns, an  access pattern does not directly  describe a path
in  the  memory  graph.  Hence  we  define  a  {\em  Canonical  Access
Pattern\/}  as  a string  restricted  to  the alphabet  $\lbrace\acar,
\acdr\rbrace$.  As a  special case,  $\bot$  is also  considered as  a
canonical access pattern.

We  define  rules  to   reduce  access  patterns  to  their  canonical
forms. For access patterns $\alpha_1$ and $\alpha_2$:
\begin{eqnarray}
    \alpha_1\bcar\alpha_2 & \rightarrow &
    \left\lbrace\begin{array}{ll}
    \alpha_1\alpha_2' & \mbox{if } \alpha_2 \equiv \acar\alpha_2' \\
    \bot & \mbox{if } \alpha_2 \equiv \acdr\alpha_2'
           \mbox{ or } \alpha_2 \equiv \epsilon
    \end{array}\right.\label{eqn:eup-red1} \\
    \alpha_1\bcdr\alpha_2 & \rightarrow &
    \left\lbrace\begin{array}{ll}
    \alpha_1\alpha_2' & \mbox{if } \alpha_2 \equiv \acdr\alpha_2' \\
    \bot & \mbox{if } \alpha_2 \equiv \acar\alpha_2'
           \mbox{ or } \alpha_2 \equiv \epsilon
    \end{array}\right.\label{eqn:eup-red2} \\
    \alpha_1\bot\alpha_2  &\rightarrow & \bot\label{eqn:eup-red3}
\end{eqnarray}
$\alpha  \rightk{k}  \alpha'$ denotes  the  reduction  of $\alpha$  to
$\alpha'$  in $k$ steps,  and $\rightstar$  denotes the  reflexive and
transitive closure of $\rightarrow$.
The concatenation ($\cdot$)  of a set  of access patterns  $\sigma_1$ with
$\sigma_2$  is  defined as  a  set  containing  concatenation of  each
element in $\sigma_1$ with each element in $\sigma_2$, i.e.

\begin{eqnarray*}
  \sigma_1\cdot\sigma_2 &\eqdef& \lbrace \alpha_1\alpha_2 \mid \alpha_1 \in
  \sigma_1, \alpha_2 \in \sigma_2 \rbrace
\end{eqnarray*}

\section{Computing Liveness Environments}
\label{sec:method}

Let $\sigma$ be the set  of access patterns specifying the liveness of
the result  of evaluating  $e$. Let \lv\  be the  liveness environment
after the evaluation of $e$.  Then the liveness environment before the
computation  of $e$  is discovered  by propagating  $\sigma$ backwards
through the body  of $e$.  This is achieved  by defining an environment
transformer for $e$, denoted \xeonly.

Since $e$  may contain applications  of primitive operations  and user
defined  functions, we  also  need transfer  functions that  propagate
$\sigma$ from  the result of  the application to the  arguments. These
functions  are denoted  by \xponly\  and \xfonly.   While  \xponly\ is
given directly  based on the  semantics of the primitive,  \xfonly\ is
inferred from the body of a function.

\subsection{Computing \xeonly}
For  an expression  $e$  at program  point $\pi$,  \xe{e}{\sigma}{\lv}
computes liveness  environment at $\pi$  where $\sigma$ is the  set of
access patterns  specifying the liveness  of the result  of evaluating
$e$ and \lv\ is the  liveness environment after the evaluation of $e$.
{Additionally, as a  side effect, the program point  $\pi$ is annotated
with the  value computed. However, we  do not show  this explicitly to
avoid clutter.}
The computation of \xe{e}{\ptexp{\pi}{\sigma}}{\lv} is as follows.
\begin{eqnarray}
  \xe{\ptexp{\pi}{\kappa}}{\sigma}{\lv} &=& \lv \\
  \xe{\ptexp{\pi}{v}}{\sigma}{\lv}  &=&   \lv \cup v.\sigma \\
  \xe{\ptexp{\pi}{(\prim\; \ptexp{\pi_1}{e_1}\;
      \ptexp{\pi_2}{e_2})}}{\sigma}{\lv}
  &=& \label{eq:prop-cons1}
  \LET\; \lv' \leftarrow \xe{\ptexp{\pi_2}{e_2}}{\xp{\prim}{2}(\sigma)}{\lv}\;
  \scmin\\
  & & \nonumber\;\;\;\; \xe{\ptexp{\pi_1}{e_1}}{\xp{\prim}{1}(\sigma)}{\lv'} \\
  & & \nonumber\mbox{where $\prim$ is one of \CONS, \scmprim} \\
  \xe{\ptexp{\pi}{(\prim\ \ptexp{\pi_1}{e_1})}}{\sigma}{\lv}
  &=& \label{eq:prop-car1}
  \xe{\ptexp{\pi_1}{e_1}}{\xp{\prim}{1}(\sigma)}{\lv} \\
  & & \nonumber\mbox{where $\prim$ is one of \CAR, \CDR, \scmnull, \scmpair} \\
  \xe{\ptexp{\pi}{(\IF\ \ptexp{\pi_1}{e_1}\; \ptexp{\pi_2}{e_2}\;
      \ptexp{\pi}{e_3})}}{\sigma}{\lv}
  &=& \LET\; \lv' \leftarrow \xe{\ptexp{\pi_3}{e_3}}{\sigma}{\lv}\;\scmin \\
  & & \nonumber\;\;\;\;\LET\; \lv'' \leftarrow \xe{\ptexp{\pi_2}{e_2}}{\sigma}{\lv}\;\scmin \\
  & & \nonumber\;\;\;\;\;\;\;\; \xe{\ptexp{\pi_1}{e_1}}{\epsilonset}{\lv' \cup \lv''} \\
  \xe{\ptexp{\pi}{(\LET\; v_1 \leftarrow \ptexp{\pi_1}{e_1}\;
      \scmin\; \ptexp{\pi_2}{e_2})}}{\sigma}{\lv}
  &=& \LET\; \lv' \leftarrow \xe{\ptexp{\pi_2}{e_2}}{\sigma}{\lv}\;\scmin \\
  & & \nonumber\;\;\;\; \xe{\ptexp{\pi_1}{e_1}}{\sigma'}{\lv' - v_1.\sigma'} \\
  & & \nonumber\mbox{where }\sigma' =
  \lbrace \alpha \mid v_1.\alpha \in \lv' \rbrace \\
  \xe{\ptexp{\pi}{(f\; \ptexp{\pi_1}{e_1}\ldots\ptexp{\pi_n}{e_n})}}
     {\sigma}{\lv} \label{eq:fun2exp}
  &=& \LET\; \lv_1 \leftarrow\xe{\ptexp{\pi_n}{e_n}}{\xf{f}{n}(\sigma)}{\lv}\;\scmin \\
  & & \nonumber\;\;\;\;\vdots \\
  & & \nonumber\;\;\;\;
     \LET\; \lv_{n-1} \leftarrow \xe{\ptexp{\pi_2}{e_2}}{\xf{f}{2}(\sigma)}{\lv_{n-2}}\;\scmin \\
  & & \nonumber\;\;\;\;\;\;\;\;\;\;\;\;\;
     \xe{\ptexp{\pi_1}{e_1}}{\xf{f}{1}(\sigma)}{\lv_{n-1}}
\end{eqnarray}

We explain the definition of \xeonly\ for the \IF\ \ expression. Since
the  value of  the conditional  expression $e_1$  is boolean  and this
value is used, the liveness access pattern with respect to which $e_1$
is  computed is  \epsilonset. Further,  since  it is  not possible  to
statically  determine whether  $e_2$ or  $e_3$ will  be  executed, the
liveness environment  with respect to  which $e_1$ is computed  is the
union of the liveness environments arising out of $e_2$ and $e_3$.

\subsection{Computing \xponly\ and \xfonly}
If  $\sigma$ is  the  set  of  access patterns  specifying  the
liveness  of the  result of  evaluating $(\prim\  e_1  \ldots e_n)$,
where $\prim$ is a primitive, then \xp{\prim}{i}($\sigma$) gives the
set of access patterns specifying the liveness of $e_i$.
We describe the transfer functions for the primitives in our language:
\CAR,  \CDR,  \CONS,  \scmnull,  \scmpair\  and  \scmprim.  The  0-ary
constructor \NIL\ does not accept any argument and is ignored.

Assume that the  live access pattern for the  result of the expression
(\CAR\ $e$) is  $\alpha$. Then, the link that is  denoted by the path
labeled $\alpha$ starting from location $\loc{(\CAR\; e)}$ can also be
denoted by a path  $\acar\alpha$ starting from location $\loc{e}$.  We
can extend the same reasoning for set of access patterns ($\sigma$) of
result, i.e.\ every  pattern in the set is prefixed  by \acar\ to give
live access pattern of $e$.  Also, since the cell corresponding to $e$
is used to  find the value of  \CAR, we need to add  $\epsilon$ to the
live access  patterns of $e$.  Reasoning about  (\CDR\ $e$) similarly,
we have
\begin{eqnarray}
\begin{array}{r@{\ }c@{\ }l@{,\hspace{.5cm}}r@{\ }c@{\ }l}
\xp{\CAR}{1}(\sigma) &=& \epsilonset \plus \acarset\cdot\sigma
\label{eqn:firstprim} 
&
\xp{\CDR}{1}(\sigma) &=& \epsilonset \plus \acdrset\cdot\sigma
\end{array}
\end{eqnarray}
As seen in Example~\ref{exmp:motiv-bar}, an access pattern of $\alpha$
for result of \CONS\ translates  to an access pattern of $\bcar\alpha$
for  its first argument,  and $\bcdr\alpha$  for its  second argument.
Since \CONS\ does  not read its arguments, the  access patterns of the
arguments do not contain $\epsilon$.
\begin{eqnarray}
\begin{array}{r@{\ }c@{\ }l@{,\hspace{.5cm}}r@{\ }c@{\ }l}
\xp{\CONS}{1}(\sigma) &=& \bcarset\cdot\sigma
&
\xp{\CONS}{2}(\sigma) &=& \bcdrset\cdot\sigma
\end{array}
\end{eqnarray}
Since the remaining  primitives read only the value  of the arguments,
the set of live access patterns of the arguments is $\epsilonset$.
\begin{eqnarray}
\begin{array}{r@{\ }c@{\ }l@{,\hspace{.5cm}}r@{\ }c@{\ }l@{,\hspace{.5cm}}r@{\ }c@{\ }l@{,\hspace{.5cm}}r@{\ }c@{\ }l}
\xp{\scmnull}{1}(\sigma) &=& \epsilonset &
\xp{\scmpair}{1}(\sigma) &=& \epsilonset &
\xp{\scmprim}{1}(\sigma) &=& \epsilonset &
\xp{\scmprim}{2}(\sigma) &=& \epsilonset\label{eqn:lastprim}
\end{array}
\end{eqnarray}

We  now consider  the transfer  function for  a user  defined function
$f$. If $\sigma$ is the set of access patterns specifying the liveness
of   the  result   of   evaluating  $(f\   e_1   \ldots  e_n)$,   then
\xf{f}{i}($\sigma$) gives  the set  of access patterns  specifying the
liveness of $e_i$.
Let $f$ be defined as:
\begin{center}
  {\sf (\DEFINE\ $(f\; v_1\; \ldots\; v_n)\;\; \pi\!:e$)}
\end{center}
Assume  that $\sigma$ is  the live  access pattern  for the  result of
$f$.  Then,  $\sigma$  is  also  the  live  access  pattern  for  $e$.
\xe{e}{\sigma}{\emptyset} computes live  access patterns for $v_i$ ($1
\leq i \leq n$) at $\pi$.  Thus, the transfer function for the
$i^{th}$ argument of $f$ is given by:
\begin{eqnarray}
  \xf{f}{i}(\sigma) &=& \lbrace \alpha \mid v_i.\alpha \in
  \xe{e}{\sigma}{\emptyset}  \rbrace \;\;\;\;\; 1\leq i\leq n
  \label{eqn:exp2fun}
\end{eqnarray}
The following  example  illustrates our analysis.
\newcommand{\pgmsigma}{\ensuremath{\lbrace\epsilon, \acdr, \acdr\acar\rbrace 
    \cup \lbrace\acdr\acar\acar\rbrace\cdot\sigma_\exit}}

\begin{example}\label{exmp:append}
  Consider  the program  in Figure~\ref{fig:mot-exmp}.   To  compute the
 transfer   functions  for   \append,  we   compute   the  environment
 transformer   \xe{e}{\sigma}{\emptyset}  in   terms  of   a  variable
 $\sigma$.  Here $e$ is the body of \append. The value of the liveness
 environment  at  each point  in  the body  of  \append\  is shown  in
 Appendix~\ref{app:compute-app}.
From the liveness information at $\pi_1$ we get:

\begin{eqnarray*}
  \xf{\fnappend}{1}(\sigma)
  &=& \epsilonset \plus \lbrace\acar\bcar\rbrace\cdot\sigma\;
  \plus\; \acdrset\cdot\xf{\fnappend}{1}(\bcdrset\cdot\sigma)\\ 
  \xf{\fnappend}{2}(\sigma)
  &=& \sigma \plus \xf{\fnappend}{2}(\bcdrset\cdot\sigma)
\end{eqnarray*}

Let  $e_\exit$  represent  the   entire  program  being  analyzed  and
$\sigma_\exit$ be  the set of access patterns  describing the liveness
of the  result.  Then, the  liveness environment at various  points in
the         $e_\exit$        can         be         computed        as
\xe{e_\exit}{\sigma_\exit}{\emptyset}.   The liveness  environments at
$\pi_{14}$ and $\pi_{12}$ are as follows:

\begin{eqnarray*}
  \lv_{\pi_{14}}&=&
  \lbrace\; \ww.(\pgmsigma)\; \rbrace \\ 
  \lv_{\pi_{12}} &=&\left\lbrace\begin{array}{c}
  \yy.\xf{\append}{1}(\pgmsigma),\\
  \zz.\xf{\append}{2}(\pgmsigma)
  \end{array}\right\rbrace \\ && \hspace*{118mm}\qed
\end{eqnarray*}
%
\end{example}

We  assume that  the entire  result of  the program  is  needed, i.e.,
$\sigma_\exit$ is $\lbrace \acar, \acdr \rbrace *$.

\section{Solving the Equations for \xfonly}
\label{sec:solve-eqns}

In general,  the equations  defining the transfer  functions \xfonly\
will be recursive.  To solve such equations, we start by guessing that
the solution will be of the form:
\begin{eqnarray}
  \xf{f}{i}(\sigma) &=& \find{f}{i} \plus \fdep{f}{i}\cdot\sigma,
\end{eqnarray}
where  \find{f}{i}  and  \fdep{f}{i}  are  sets of  strings  over  the
alphabet $\lbrace  \acar, \acdr,\bcar, \bcdr  \rbrace$.  The intuition
behind this form  of solution is as follows: The  function $f$ can use
its argument  locally and/or  copy a  part of it  to the  return value
being  computed. \find{f}{i} is  the live  access pattern  of $i^{th}$
argument due to  local use in $f$.  \fdep{f}{i} is  a sort of selector
that  selects  the liveness  pattern  corresponding  to the  $i^{th}$
argument  of $f$  from $\sigma$,  the liveness  pattern of  the return
value.

If  we substitute  the  guessed  form of  \xf{f}{i}  in the  equations
describing it and  equate the terms containing $\sigma$  and the terms
without   $\sigma$,  we   get  the   equations  for   \find{f}{i}  and
\fdep{f}{i}. This is illustrated in the following example.

\begin{example}\label{exmp:decompose-eqn}
  Consider   the   equation   for   \xf{\fnappend}{1}($\sigma$)   from
  Example~\ref{exmp:append}:
\begin{eqnarray*}
  \xf{\fnappend}{1}(\sigma)
  &=& \epsilonset \plus \lbrace\acar\bcar\rbrace\cdot\sigma \plus
  \acdrset\cdot\xf{\fnappend}{1}(\bcdrset\cdot\sigma)
\end{eqnarray*}
Decomposing both sides of the equation, and rearranging the RHS gives:
\begin{eqnarray*}
  \find{\fnappend}{1} \plus
  \fdep{\fnappend}{1}\cdot\sigma
  &=&
  \epsilonset \plus \lbrace\acar\bcar\rbrace\cdot\sigma \plus
  \acdrset\cdot(\find{\fnappend}{1} \plus\;
  \fdep{\fnappend}{1}\cdot\bcdrset\cdot\sigma) \\
  &=&
  \epsilonset \plus \acdrset\cdot\find{\mbox{\footnotesize
  \append}}{1} \plus \lbrace\acar\bcar\rbrace\cdot\sigma \plus
  \acdrset\cdot\fdep{\fnappend}{1}\cdot\bcdrset\cdot\sigma
\end{eqnarray*}
Separating the parts that are $\sigma$ dependent and the parts that
are $\sigma$ independent, and   equating  them separately, we get:
\begin{eqnarray*}
  \find{\fnappend}{1} &=& \epsilonset
  \plus \acdrset\cdot\find{\fnappend}{1} \\
  \fdep{\fnappend}{1}\cdot\sigma &=& \lbrace\acar\bcar\rbrace\cdot\sigma
  \plus \acdrset\cdot\fdep{\fnappend}{1}\cdot\bcdrset\sigma \\
  &=& (\lbrace\acar\bcar\rbrace
  \plus \acdrset\cdot\fdep{\fnappend}{1}\cdot\bcdrset)\cdot\sigma
\end{eqnarray*}
As the equations hold for any general $\sigma$, we can simplify them to:
\begin{eqnarray*}
  \begin{array}{rcllrcl}
  \find{\fnappend}{1} &=& \epsilonset
  \plus \acdrset\cdot\find{\fnappend}{1} 
  & \mbox{ and } &  \fdep{\fnappend}{1} &=& \lbrace\acar\bcar\rbrace 
  \plus \acdrset\cdot\fdep{\fnappend}{1}\cdot\bcdrset
  \end{array}
\end{eqnarray*}
Similarly, from the equation describing \xf{\fnappend}{2}($\sigma$),
we get:
\begin{eqnarray*}
  \begin{array}{r@{\ }c@{\ }llr@{\ }c@{\ }l}
  \find{\fnappend}{2} &=& \find{\fnappend}{2}
  & \mbox{ and } &
  \fdep{\fnappend}{2} &=& \epsilonset
  \plus \fdep{\fnappend}{2}\cdot\bcdrset
  \end{array}
\end{eqnarray*}
The liveness environment at $\pi_{12}$ and $\pi_{14}$ in terms
\find{\append}{} and \fdep{\append}{} are:
\begin{eqnarray*}
  \lv_{\pi_{14}}&=&
  \lbrace\; \ww.\lbrace\acdr\acar\acar\rbrace\cdot\sigma_\exit\;\rbrace\\
  \lv_{\pi_{12}} &=&\left\lbrace\begin{array}{@{}c@{}}
  \yy.(\find{\append}{1} \;\cup\; \fdep{\append}{1}\cdot(\pgmsigma)),\\
  \zz.(\find{\append}{2} \;\cup\; \fdep{\append}{2}\cdot(\pgmsigma))
  \end{array}\right\rbrace
\end{eqnarray*}
Solving for \find{\append}{} and \fdep{\append}{} gives us the desired
liveness environments at these program points.
\qed\end{example}
\subsection{Representing Liveness by Context Free Grammars}
\label{sec:cfg-eqn}

\newcommand{\var}[1]{\mbox{$\langle$#1$\rangle$}}

The values of \find{}{} and \fdep{}{} variables of a transfer function
are sets  of strings over  the alphabet $\lbrace \acar,  \acdr, \bcar,
\bcdr \rbrace$.  We use context  free grammars (CFG) to describe these
sets.
The set of terminal symbols of the CFG is $\lbrace\acar, \acdr, \bcar,
\bcdr\rbrace$.  Non-terminals and  associated rules are constructed as
illustrated             in             Examples~\ref{exmp:cfg-eqn-app}
and~\ref{exmp:cfg-eqn-full}.

\begin{example}\label{exmp:cfg-eqn-app}
  Consider         the        following         constraint        from
  Example~\ref{exmp:decompose-eqn}:
\begin{eqnarray*}
  \find{\fnappend}{1} &=& \epsilonset
  \plus \acdrset\cdot\find{\fnappend}{1}
\end{eqnarray*}
 We  add non-terminal  \var{\find{\fnappend}{1}}  and the 
 productions with right hand sides directly derived from the constraints:
\begin{eqnarray*}
  \var{\find{\fnappend}{1}} &\rightarrow& \epsilon \mid
  \acdr\var{\find{\fnappend}{1}}
\end{eqnarray*}
The productions generated from other constraints of
Example~\ref{exmp:decompose-eqn} are:
\begin{eqnarray*}
  \var{\fdep{\fnappend}{1}} &\rightarrow& \acar\bcar \mid
  \acdr\var{\fdep{\fnappend}{1}}\bcdr\\
  \var{\find{\fnappend}{2}} &\rightarrow& \var{\find{\fnappend}{2}} \\
  \var{\fdep{\fnappend}{2}} &\rightarrow& \epsilon \mid
   \var{\fdep{\fnappend}{2}}\bcdr
\end{eqnarray*}
These productions describe the transfer functions of \append.
\qed
\end{example}

The liveness environment at each program point can be represented as a
CFG with  a start symbol for  every variable.  To do  so, the analysis
starts with \var{$S_\exit$},  the non-terminal describing the liveness
pattern of  the result of the program,  $\sigma_\exit$.
The  productions  for \var{$S_\exit$} are:

\begin{eqnarray*}
  \var{$S_\exit$} &\rightarrow& \epsilon \mid \acar \var{$S_\exit$}
  \mid\acdr \var{$S_\exit$}
\end{eqnarray*}

\begin{example}\label{exmp:cfg-eqn-full}
  Let {$S_{\pi}^{\sf v}$}  denote the non-terminal generating liveness
  access patterns associated with a  variable {\sf v} at program point
  $\pi$.  For the program of Figure~\ref{fig:mot-exmp}:

  \begin{eqnarray*}
    \var{$S_{\pi_{14}}^{\ww}$} &\rightarrow&
    \epsilon \mid \acdr \mid \acdr\acar \mid \acdr\acar\acar\var{$S_\exit$}\\
    \var{$S_{\pi_{12}}^{\zz}$} &\rightarrow&
    \var{\find{\append}{2}} \mid
    \var{\fdep{\append}{2}} \mid 
    \var{\fdep{\append}{2}}\acdr \mid
    \var{\fdep{\append}{2}}\acdr\acar \mid
    \var{\fdep{\append}{2}}\acdr\acar\acar \var{$S_\exit$} \\
    \var{$S_{\pi_{12}}^{\yy}$} &\rightarrow&
    \var{\find{\append}{1}} \mid 
    \var{\fdep{\append}{1}} \mid
    \var{\fdep{\append}{1}}\acdr \mid
    \var{\fdep{\append}{1}}\acdr\acar \mid
    \var{\fdep{\append}{1}}\acdr\acar\acar \var{$S_\exit$}
    \hspace*{23mm}\qed
  \end{eqnarray*}  
\end{example}

The access patterns in the  access paths used for nullification are in
canonical form but the access patterns described by the CFGs resulting
out  of our  analysis are  not. It  is not  obvious how  to  check the
membership of a canonical access  pattern in such CFGs.  To solve this
problem, we need  equivalent CFGs such that if  $\alpha$ belongs to an
original  CFG  and $\alpha  \rightstar  \beta$,  where  $\beta$ is  in
canonical form, then $\beta$ belongs to the corresponding new CFG.
Directly         converting         the        reduction         rules
(Equations~(\ref{eqn:eup-red1},                     \ref{eqn:eup-red2},
\ref{eqn:eup-red3}))  into productions  and adding  it to  the grammar
results   in  {\em  unrestricted   grammar}~\cite{hopcraft90toc}.   To
simplify    the   problem,   we    approximate   original    CFGs   by
non-deterministic  finite  automata (NFAs)  and  eliminate \bcar\  and
\bcdr\ from the NFAs.

\subsection{Approximating CFGs using NFAs}
\label{sec:nfa-eqn}

The conversion of a CFG $\gram$ to an approximate NFA $\nfa$ should be
safe in that  the language accepted by $\nfa$ should  be a superset of
the language  accepted by $\gram$.  We use the algorithm  described by
Mohri and  Nederhof~\cite{mohri00regular}. The algorithm  transforms a
CFG to a  restricted form called {\em strongly  regular} CFG which can
be converted easily to a finite automaton.

\newcommand{\nfaNN}[1]{%
    \psset{unit=1mm}
  \begin{pspicture}(0,0)(15,13)
    \putnode{n0}{origin}{0}{3}{}
    \putnode{nn}{n0}{10}{0}{
      \pscirclebox[framesep=2,doubleline=true]{\mbox{ }}}
    \ncline[nodesepB=-1]{->}{n0}{nn}\Aput[1]{start}
    \nccurve[angleA=110,angleB=70,ncurv=5,nodesepB=0]{->}{nn}{nn}\Aput[.1]{#1}
  \end{pspicture}
}

\begin{example}\label{exmp:nfa-eqn}
We  show  the  approximate  NFAs  for each  of  the  non-terminals  in
Example~\ref{exmp:cfg-eqn-app} and Example~\ref{exmp:cfg-eqn-full}.
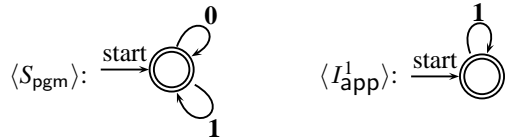
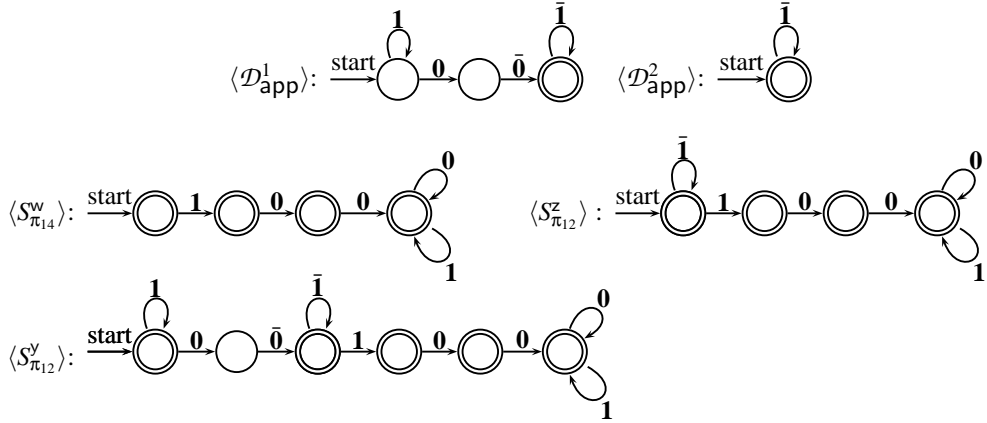
\begin{center}
\scalebox{.9}{
\begin{tabular}{@{}l@{\ \ }ll@{\ \ }l@{}}
  \var{$S_\exit$}:&    
  \raisebox{-5mm}{\scalebox{1}{\psset{unit=1mm}
  \begin{pspicture}(0,0)(28,18)
    \putnode{n0}{origin}{0}{7}{}
    \putnode{na}{n0}{10}{0}{
      \pscirclebox[framesep=2,doubleline=true]{\mbox{ }}}
    \ncline[nodesepB=-1]{->}{n0}{na}\Aput[1]{start}
    \nccurve[angleA=70,angleB=30,ncurv=5,nodesepB=-.5]{->}{na}{na}\Aput[.1]{\acar}
    \nccurve[angleA=-30,angleB=-70,ncurv=5]{->}{na}{na}\Aput[.2]{\acdr}
  \end{pspicture}}}
  &
  \var{\find{\fnappend}{1}}:&
  \raisebox{-2mm}{\scalebox{1}{\nfaNN{\acdr}}}
  \\
\end{tabular}}
\end{center}
\begin{center}
\scalebox{.9}{
\begin{tabular}{@{}l@{\ \ }ll@{\ \ }l@{}}
  \var{\fdep{\fnappend}{1}}:&  
  \raisebox{-3mm}{\scalebox{1}{\psset{unit=1mm}
  \begin{pspicture}(0,0)(38,15)
    \putnode{n0}{origin}{0}{4}{}
    \putnode{na}{n0}{10}{0}{\pscirclebox[framesep=2.4]{\mbox{ }}}
    \ncline[nodesepB=0]{->}{n0}{na}\Aput[1]{start}
    \nccurve[angleA=110,angleB=70,ncurv=5]{->}{na}{na}\Aput[.1]{\acdr}
    \putnode{nb}{na}{12}{0}{\pscirclebox[framesep=2.4]{\mbox{ }}}
    \putnode{nc}{nb}{12}{0}{\pscirclebox[framesep=2,doubleline=true]{\mbox{ }}}
    \nccurve[angleA=110,angleB=70,ncurv=5]{->}{nc}{nc}\Aput[.2]{\bcdr}

    \ncline[nodesep=0]{->}{na}{nb}\Aput[.1]{\acar}
    \ncline[nodesep=0]{->}{nb}{nc}\Aput[.1]{\bcar}
  \end{pspicture}}}
  &
  \var{\fdep{\fnappend}{2}}:& \raisebox{-2mm}{\scalebox{1}{\nfaNN{\bcdr}}}
  \\
\end{tabular}}
\end{center}
\begin{center}
\scalebox{.9}{
\begin{tabular}{@{}l@{\ \ }ll@{\ \ }l@{}}
  \raisebox{8mm}{\var{$S_{\pi_{14}}^{\ww}$}:} &
  \multicolumn{1}{@{}l@{}}{\scalebox{1}{\psset{unit=1mm}
  \begin{pspicture}(0,0)(63,22)
    \putnode{n0}{origin}{0}{9}{}
    \putnode{na}{n0}{10}{0}{\pscirclebox[framesep=2,doubleline=true]{\mbox{ }}}
    \ncline[nodesepB=0]{->}{n0}{na}\Aput[1]{start}
    \putnode{nb}{na}{12}{0}{\pscirclebox[framesep=2,doubleline=true]{\mbox{ }}}
    \putnode{nc}{nb}{12}{0}{\pscirclebox[framesep=2,doubleline=true]{\mbox{ }}}

    \ncline[nodesep=0]{->}{na}{nb}\Aput[.1]{\acdr}
    \ncline[nodesep=0]{->}{nb}{nc}\Aput[.1]{\acar}
    \putnode{nd}{nc}{13}{0}{
      \pscirclebox[framesep=2,doubleline=true]{\mbox{ }}}
    \ncline[nodesepB=-1]{->}{nc}{nd}\Aput[.2]{\acar}
    \nccurve[angleA=70,angleB=30,ncurv=5,nodesepB=-.5]{->}{nd}{nd}\Aput[.1]{\acar}
    \nccurve[angleA=-30,angleB=-70,ncurv=5]{->}{nd}{nd}\Aput[.2]{\acdr}
  \end{pspicture}}} 
   &
  \raisebox{8mm}{\var{$S_{\pi_{12}}^{\zz}$} :} &
  \multicolumn{1}{@{}l@{}}{\scalebox{1}{\psset{unit=1mm}
  \begin{pspicture}(0,0)(63,22)
    \putnode{mb}{origin}{0}{9}{}
    \putnode{ma}{mb}{10}{0}{\pscirclebox[framesep=2,doubleline=true]{\mbox{ }}}
    \ncline[nodesepB=0]{->}{mb}{ma}\Aput[1]{start}
    \nccurve[angleA=110,angleB=70,ncurv=5,nodesepB=0]{->}{ma}{ma}\Aput[.1]{\bcdr}
    \putnode{na}{ma}{0}{0}{\pscirclebox[framesep=2,doubleline=true]{\mbox{ }}}
    \putnode{nb}{na}{12}{0}{\pscirclebox[framesep=2,doubleline=true]{\mbox{ }}}
    \putnode{nc}{nb}{12}{0}{\pscirclebox[framesep=2,doubleline=true]{\mbox{ }}}

    \ncline[nodesep=0]{->}{na}{nb}\Aput[.1]{\acdr}
    \ncline[nodesep=0]{->}{nb}{nc}\Aput[.1]{\acar}
    \putnode{nd}{nc}{13}{0}{
      \pscirclebox[framesep=2,doubleline=true]{\mbox{ }}}
    \ncline[nodesepB=-1]{->}{nc}{nd}\Aput[.2]{\acar}
    \nccurve[angleA=70,angleB=30,ncurv=5,nodesepB=-.5]{->}{nd}{nd}\Aput[.1]{\acar}
    \nccurve[angleA=-30,angleB=-70,ncurv=5]{->}{nd}{nd}\Aput[.2]{\acdr}
  \end{pspicture}}}
  \\
  \raisebox{5mm}{\var{$S_{\pi_{12}}^{\yy}$}:} &
  \multicolumn{3}{@{}l@{}}{\scalebox{1}{\psset{unit=1mm}
  \begin{pspicture}(0,0)(85,17)
    \putnode{n0}{origin}{0}{7}{}
    \putnode{na}{n0}{10}{0}{\pscirclebox[framesep=2,doubleline=true]{\mbox{ }}}
    \ncline[nodesepB=0]{->}{n0}{na}\Aput[1]{start}
    \nccurve[angleA=110,angleB=70,ncurv=5]{->}{na}{na}\Aput[.1]{\acdr}
    \putnode{nb}{na}{12}{0}{\pscirclebox[framesep=2.4]{\mbox{ }}}
    \putnode{nc}{nb}{12}{0}{\pscirclebox[framesep=2,doubleline=true]{\mbox{ }}}
    \nccurve[angleA=110,angleB=70,ncurv=5]{->}{nc}{nc}\Aput[.2]{\bcdr}
    \ncline[nodesep=0]{->}{na}{nb}\Aput[.1]{\acar}
    \ncline[nodesep=0]{->}{nb}{nc}\Aput[.1]{\bcar}
    
    \putnode{ma}{nc}{0}{0}{\pscirclebox[framesep=2,doubleline=true]{\mbox{ }}}
    \ncline[nodesepB=0]{->}{n0}{na}\Aput[1]{start}
    \putnode{mb}{ma}{12}{0}{\pscirclebox[framesep=2,doubleline=true]{\mbox{ }}}
    \putnode{mc}{mb}{12}{0}{\pscirclebox[framesep=2,doubleline=true]{\mbox{ }}}

    \ncline[nodesep=0]{->}{ma}{mb}\Aput[.1]{\acdr}
    \ncline[nodesep=0]{->}{mb}{mc}\Aput[.1]{\acar}
    \putnode{md}{mc}{12}{0}{
      \pscirclebox[framesep=2,doubleline=true]{\mbox{ }}}
    \ncline[nodesepB=-1]{->}{mc}{md}\Aput[.2]{\acar}
    \nccurve[angleA=70,angleB=30,ncurv=5,nodesepB=-.5]{->}{md}{md}\Aput[.1]{\acar}
    \nccurve[angleA=-30,angleB=-70,ncurv=5]{->}{md}{md}\Aput[.2]{\acdr}
  \end{pspicture}}}
\end{tabular}
}
\end{center}
\noindent     Note    that     there    is     no     automaton    for
\var{\find{\fnappend}{2}}. This  is because the least  solution of the
equation          \mbox{$\var{\find{\fnappend}{2}}         \rightarrow
\var{\find{\fnappend}{2}}$}   is  $\emptyset$.   Also,   the  language
accepted by the automaton for \fdep{\fnappend}{1} is approximate as it
does not ensure that there is  an equal number of \acdr\ and \bcdr\ in
the strings generated by rules for \var{\fdep{\fnappend}{1}}.
\qed\end{example} 

\subsection{Eliminating \bcar\ and \bcdr\ from NFA}
\label{sec:nfa-elim}

We  now describe how  to convert  an NFA  with transitions  on symbols
\bcar\  and \bcdr\  to an  equivalent NFA  without any  transitions on
these symbols.

\noindent  {\bf  Input:}  An NFA  \nfabar\  with  underlying  alphabet
  $\lbrace  \acar,  \acdr, \bcar,  \bcdr\rbrace$  accepting  a set  of
  access patterns\\
  {\bf Output:} An NFA \nfa\ with underlying  alphabet $\lbrace \acar,
  \acdr\rbrace$  accepting  the  equivalent  set of  canonical  access
  patterns\\
  {\bf Steps:}
  \begin{center}
    \scalebox{.99}{
  \begin{uprogram}
    \UFL \ \ $i \leftarrow 0$ 
    \UNL{0} $\nfa_0 \leftarrow$ Equivalent NFA of
    \nfabar\  without $\epsilon$-moves \cite{hopcraft90toc}
    \UNL{0} {\bf do} \{
    \UNL{1} $\nfa'_{i+1} \leftarrow \nfa_i$
    \UNL{1} foreach state $q$ in $\nfa_i$ such that $q$ has an incoming
    edge from $q'$
    \UNL{1}   with label $\bcar$ and outgoing edge to $q''$ with label
    $\acar$ \{\ \hfill
    \UNL{2} /$\star$ bypass  $\bcar\acar$ using $\epsilon\ \star$/
    \UNL{2} add an edge in $\nfa'_{i+1}$ from $q'$ to $q''$ with label
    $\epsilon$.
    \UNL{1} \}

    \UNL{1} foreach state $q$ in $\nfa_i$ such that $q$ has an incoming
    edge from $q'$
    \UNL{1}   with label $\bcdr$ and outgoing edge to $q''$ with label
    $\acdr$ \{\ \hfill 
    \UNL{2} /$\star$ bypass  $\bcdr\acdr$ using $\epsilon\ \star$/

    \UNL{2} add an edge in $\nfa'_{i+1}$ from $q'$ to $q''$ with label
    $\epsilon$.
    \UNL{1} \}

    \UNL{1} $\nfa_{i+1} \leftarrow$ Equivalent NFA of $\nfa'_{i+1}$ without
    $\epsilon$-moves
    
    \UNL{1} $i \leftarrow i+1$
    \UNL{0} \} {\bf while} ($\nfa_{i} \not= \nfa_{i-1}$)

    \UNL{0} $\nfa \leftarrow \nfa_i$
    \UNL{0} delete all edges with label \bcar\ or \bcdr\ in \nfa.
  \end{uprogram}}
  \end{center}

\noindent  The  algorithm repeatedly  introduces  $\epsilon$ edges  to
bypass a pair of  consecutive edges labeled \bcar\acar\ or \bcdr\acdr.
The  process is continued  till a  fixed point  is reached.   When the
fixed  point is  reached,  the resulting  NFA  contains the  canonical
access  patterns  corresponding to  all  the  access  patterns in  the
original NFA. The access patterns not in canonical form are deleted by
removing edges labeled  \bcar\ and \bcdr.  Note that  by our reduction
rules  if $\alpha$  is accepted  by $\nfabar$  and  $\alpha \rightstar
\bot$,  then $\bot$  should  be accepted  by  $\nfa$, However,  $\nfa$
returned  by our  algorithm does  not accept  $\bot$.  This  is  not a
problem because  the access patterns  which are tested  for membership
against $\nfa$ do not include $\bot$ as well.

\begin{example}\label{exmp:elim-01}
We  show the elimination  of \bcar\  and \bcdr\  for the  automata for
\var{$S_{\pi_{12}}^{\yy}$}    and   \var{$S_{\pi_{12}}^{\zz}$}.    The
automaton for \var{$S_{\pi_{14}}^{\ww}$}  remains unchanged as it does
not  contain transitions  on \bcar\  and \bcdr.   The automata  at the
termination of the loop in the algorithm are:
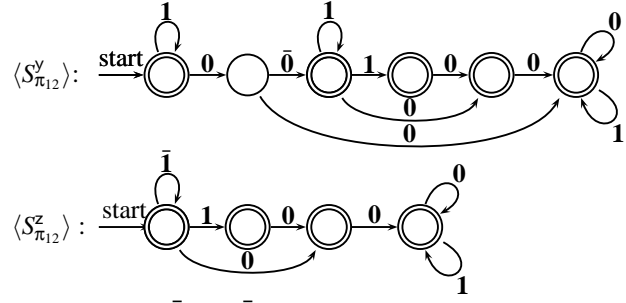
\begin{center}
\scalebox{.9}{
\begin{tabular}{@{}l@{\ \ }ll@{\ \ }l@{}}
  \raisebox{10mm}{\var{$S_{\pi_{12}}^{\yy}$}:} &
  \multicolumn{3}{@{}l@{}}{\scalebox{1}{\psset{unit=1mm}
  \begin{pspicture}(0,0)(85,23)
    \putnode{n0}{origin}{0}{11}{}
    \putnode{na}{n0}{10}{0}{\pscirclebox[framesep=2,doubleline=true]{\mbox{ }}}
    \ncline[nodesepB=0]{->}{n0}{na}\Aput[1]{start}
    \nccurve[angleA=110,angleB=70,ncurv=5]{->}{na}{na}\Aput[.1]{\acdr}
    \putnode{nb}{na}{12}{0}{\pscirclebox[framesep=2.4]{\mbox{ }}}
    \putnode{nc}{nb}{12}{0}{\pscirclebox[framesep=2,doubleline=true]{\mbox{ }}}
    \nccurve[angleA=110,angleB=70,ncurv=5]{->}{nc}{nc}\Aput[.2]{\bcdr}
    \ncline[nodesep=0]{->}{na}{nb}\Aput[.1]{\acar}
    \ncline[nodesep=0]{->}{nb}{nc}\Aput[.1]{\bcar}
    
    \putnode{ma}{nc}{0}{0}{\pscirclebox[framesep=2,doubleline=true]{\mbox{ }}}
    \ncline[nodesepB=0]{->}{n0}{na}\Aput[1]{start}
    \putnode{mb}{ma}{12}{0}{\pscirclebox[framesep=2,doubleline=true]{\mbox{ }}}
    \putnode{mc}{mb}{12}{0}{\pscirclebox[framesep=2,doubleline=true]{\mbox{ }}}

    \ncline[nodesep=0]{->}{ma}{mb}\Aput[.1]{\acdr}
    \ncline[nodesep=0]{->}{mb}{mc}\Aput[.1]{\acar}
    \putnode{md}{mc}{12}{0}{
      \pscirclebox[framesep=2,doubleline=true]{\mbox{ }}}
    \ncline[nodesepB=-1]{->}{mc}{md}\Aput[.2]{\acar}
    \nccurve[angleA=70,angleB=30,ncurv=5,nodesepB=-.5]{->}{md}{md}\Aput[.1]{\acar}
    \nccurve[angleA=-30,angleB=-70,ncurv=5]{->}{md}{md}\Aput[.2]{\acdr}
    \nccurve[angleA=-60,angleB=-120,ncurv=.5]{->}{nc}{mc}\Aput[.2]{\acar}
    \nccurve[angleA=-60,angleB=-120,ncurv=.5]{->}{nb}{md}\Aput[.2]{\acar}
  \end{pspicture}}}\\
  \raisebox{6mm}{\var{$S_{\pi_{12}}^{\zz}$} :} &
  \multicolumn{3}{@{}l@{}}{\scalebox{1}{\psset{unit=1mm}
  \begin{pspicture}(0,0)(63,17)
    \putnode{mb}{origin}{0}{7}{}
    \putnode{ma}{mb}{10}{0}{\pscirclebox[framesep=2,doubleline=true]{\mbox{ }}}
    \ncline[nodesepB=-1]{->}{mb}{ma}\Aput[1]{start}
    \nccurve[angleA=110,angleB=70,ncurv=5,nodesepB=0]{->}{ma}{ma}\Aput[.1]{\bcdr}
    \putnode{na}{ma}{0}{0}{\pscirclebox[framesep=2,doubleline=true]{\mbox{ }}}
    \putnode{nb}{na}{12}{0}{\pscirclebox[framesep=2,doubleline=true]{\mbox{ }}}
    \putnode{nc}{nb}{12}{0}{\pscirclebox[framesep=2,doubleline=true]{\mbox{ }}}

    \ncline[nodesep=0]{->}{na}{nb}\Aput[.1]{\acdr}
    \ncline[nodesep=0]{->}{nb}{nc}\Aput[.1]{\acar}
    \putnode{nd}{nc}{13}{0}{
      \pscirclebox[framesep=2,doubleline=true]{\mbox{ }}}
    \ncline[nodesepB=-1]{->}{nc}{nd}\Aput[.2]{\acar}
    \nccurve[angleA=70,angleB=30,ncurv=5,nodesepB=-.5]{->}{nd}{nd}\Aput[.1]{\acar}
    \nccurve[angleA=-30,angleB=-70,ncurv=5]{->}{nd}{nd}\Aput[.2]{\acdr}
    \nccurve[angleA=-60,angleB=-120,ncurv=.5]{->}{ma}{nc}\Aput[.2]{\acar}
  \end{pspicture}}}  
\end{tabular}
}
\end{center}
\noindent Eliminating the edges labeled \bcar\ and \bcdr, and removing
the dead states gives:
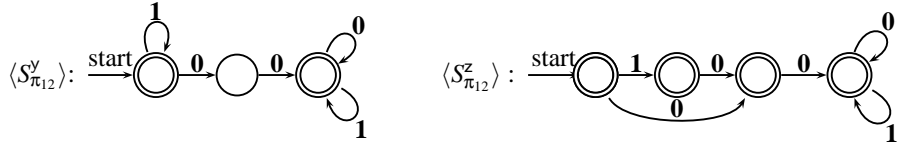
\begin{center}
\scalebox{.9}{
\begin{tabular}{@{}l@{\ \ }ll@{\ \ }ll@{\ \ }ll@{\ \ }l@{}}
  \raisebox{6mm}{\var{$S_{\pi_{12}}^{\yy}$}:} &
  \multicolumn{3}{@{}l@{}}{\scalebox{1}{\psset{unit=1mm}
  \begin{pspicture}(0,0)(50,17)
    \putnode{n0}{origin}{0}{7}{}
    \putnode{na}{n0}{10}{0}{\pscirclebox[framesep=2,doubleline=true]{\mbox{ }}}
    \ncline[nodesepB=0]{->}{n0}{na}\Aput[1]{start}
    \nccurve[angleA=110,angleB=70,ncurv=5]{->}{na}{na}\Aput[.1]{\acdr}
    \putnode{nb}{na}{12}{0}{\pscirclebox[framesep=2.4]{\mbox{ }}}
    \ncline[nodesep=0]{->}{na}{nb}\Aput[.1]{\acar}
    

    \putnode{md}{nb}{12}{0}{\pscirclebox[framesep=2,doubleline=true]{\mbox{ }}}
    \nccurve[angleA=70,angleB=30,ncurv=5,nodesepB=-.5]{->}{md}{md}\Aput[.1]{\acar}
    \nccurve[angleA=-30,angleB=-70,ncurv=5]{->}{md}{md}\Aput[.2]{\acdr}
    \ncline{->}{nb}{md}\Aput[.2]{\acar}
  \end{pspicture}}}  
  &
  \raisebox{6mm}{\var{$S_{\pi_{12}}^{\zz}$} :} &
  \multicolumn{3}{@{}l@{}}{\scalebox{1}{\psset{unit=1mm}
  \begin{pspicture}(0,0)(63,17)
    \putnode{mb}{origin}{0}{7}{}
    \putnode{ma}{mb}{10}{0}{\pscirclebox[framesep=2,doubleline=true]{\mbox{ }}}
    \ncline[nodesepB=-1]{->}{mb}{ma}\Aput[1]{start}
    \putnode{na}{ma}{0}{0}{\pscirclebox[framesep=2,doubleline=true]{\mbox{ }}}
    \putnode{nb}{na}{12}{0}{\pscirclebox[framesep=2,doubleline=true]{\mbox{ }}}
    \putnode{nc}{nb}{12}{0}{\pscirclebox[framesep=2,doubleline=true]{\mbox{ }}}

    \ncline[nodesep=0]{->}{na}{nb}\Aput[.1]{\acdr}
    \ncline[nodesep=0]{->}{nb}{nc}\Aput[.1]{\acar}
    \putnode{nd}{nc}{13}{0}{
      \pscirclebox[framesep=2,doubleline=true]{\mbox{ }}}
    \ncline[nodesepB=-1]{->}{nc}{nd}\Aput[.2]{\acar}
    \nccurve[angleA=70,angleB=30,ncurv=5,nodesepB=-.5]{->}{nd}{nd}\Aput[.1]{\acar}
    \nccurve[angleA=-30,angleB=-70,ncurv=5]{->}{nd}{nd}\Aput[.2]{\acdr}
    \nccurve[angleA=-60,angleB=-120,ncurv=.5]{->}{ma}{nc}\Aput[.2]{\acar}
  \end{pspicture}}}  
\end{tabular}
}
\end{center}
The  language accepted  by these  automata represent  the  live access
paths corresponding to \yy\ and \zz\ at $\pi_{12}$.
\qed\end{example}

We now prove the termination and correctness of our algorithm.

\newcommand{\mylongrightarrow}{
  \ \pnode{p0}\hspace{10.5mm}\pnode{p1}\ %
  \ncline{->}{p0}{p1}
}

\newcommand{\rightarroweps}{\mbox{\raisebox{1.2mm}{\mylongrightarrow
   \Aput[.1]{\footnotesize addition}
   \Bput[.1]{\footnotesize of $\epsilon$-edges}
}}}
\newcommand{\rightarrownoeps}{\mbox{\raisebox{1.2mm}{\mylongrightarrow
   \Aput[.1]{\footnotesize deletion}
   \Bput[.1]{\footnotesize of $\epsilon$-edges}
}}}

\newcommand{\rightarrownobar}{\mbox{\raisebox{1.2mm}{\mylongrightarrow
   \Aput[.1]{\footnotesize deletion of }
   \Bput[.1]{\footnotesize \bcar, \bcdr\ edges}
}}}

\subsubsection*{Termination}
Termination  of  the  algorithm  follows  from  the  fact  that  every
iteration of {\bf do-while} loop adds  new edges to the NFA, while old
edges are not  deleted.  Since no new states are added  to NFA, only a
fixed number of  edges can be added before we reach  a fix point.

\subsubsection*{Correctness}
The sequence of obtaining \nfa\ from \nfabar\
can be viewed as follows, with $\nfa_m$ denoting the NFA at the
termination of while loop:
$$    \nfabar   \rightarrownoeps    \nfa_0    \rightarroweps   \nfa'_1
\rightarrownoeps  \nfa_1\rightarroweps  \cdots \rightarroweps  \nfa'_i
\rightarrownoeps  \nfa_i  \cdots  \rightarrownoeps  \nfa_m$$ 
$$\nfa_m \rightarrownobar \nfa$$
Then, the languages accepted by these NFAs have the following
relation:
$$ L(\nfabar)  = L(\nfa_0) \subseteq L(\nfa'_1)  = L(\nfa_1) \subseteq
\cdots \subseteq L(\nfa'_i) = L(\nfa_i) \subseteq \cdots = L(\nfa_m)$$
$$ L(\nfa) \subseteq L(\nfa_m) $$

We first prove that the addition of $\epsilon$-edges in the while loop
does not add any new information, i.e.\ any access pattern accepted by
the NFA after the addition of $\epsilon$-edges is a reduced version of
some  access  pattern existing  in  the  NFA  before the  addition  of
$\epsilon$-edges.

\begin{lemma}\label{lemma:no-new}
  for $i > 0$, if $\alpha \in L(\nfa_i)$ then there exists $\alpha'
  \in L(\nfa_{i-1})$ such that $\alpha' \rightstar \alpha$.
\end{lemma}
\begin{proof}
  As $ L(\nfa_i) = L(\nfa'_i)$, we have $\alpha \in L(\nfa'_i)$. Only
  difference between $\nfa'_i$ and $\nfa_{i-1}$ is that $\nfa'_i$
  contains some extra $\epsilon$-edges. Thus, any $\epsilon$-edge free
  path in $\nfa'_i$ is also in $\nfa_{i-1}$. Consider a path $p$ in
  $\nfa'_i$ that accepts $\alpha$. Assume the number of $\epsilon$
  edges in $p$ is $k$. The proof is by induction on $k$.\\
  \noindent{({\em BASE})} $k = 0$, i.e.\ $p$ does not contains any
    $\epsilon$-edge: As the path $p$ is $\epsilon$-edge free, it must
    be present in $\nfa_{i-1}$. Thus, $\nfa_{i-1}$ also accepts
    $\alpha$. $\alpha \rightstar \alpha$.\\
  \noindent{({\em HYPOTHESIS})} For any $\alpha \in L(\nfa_i) $ with
    accepting path $p$ having less than $k$ $\epsilon$-edges there
    exists $\alpha' \in L(\nfa_{i-1})$ such that $\alpha' \rightstar
    \alpha$. \\
  \noindent{({\em  INDUCTION})}   $p$  contains  $k$  $\epsilon$-edges
    $e_1,\ldots,e_k$: Assume  $e_1$ connects states $q'$  and $q''$ in
    $\nfa'_i$. By construction, there  exists a state $q$ in $\nfa'_i$
    such that  there is  an edge  $e'_1$ from $q'$  to $q$  with label
    \bcar(\bcdr)  and an  edge $e''_1$  from $q$  to $q''$  with label
    \acar(\acdr) in $\nfa'_i$. Replace  $e_1$ by $e'_1e''_1$ in $p$ to
    get a new  path $p''$ in $\nfa'_i$.  Let  $\alpha''$ be the access
    pattern accepted by  $p''$. Clearly, $\alpha'' \rightk{1} \alpha$.
    Since $p''$ has $k-1$  $\epsilon$-edges, $\alpha''$ is accepted by
    $\nfa'_i$   along  a  path   ($p''$)  that   has  less   than  $k$
    $\epsilon$-edges.  By  induction hypothesis, we  have $\alpha' \in
    L(\nfa_{i-1})$ such  that $\alpha'\rightstar\alpha''$.  This along
    with $\alpha'' \rightk{1} \alpha$ gives $\alpha'\rightstar\alpha$.
\end{proof}

\begin{corollary}\label{corr:no-new}
  for each $\alpha \in L(\nfa_m)$, there exists $\alpha' \in
  L(\nfabar)$ such that $\alpha' \rightstar \alpha$.
\end{corollary}
\begin{proof}
  The    proof    is    by     induction    on    $m$,    and    using
  Lemma~\ref{lemma:no-new}.
\end{proof} 

The following lemma  shows that the the language  accepted by $\nfa_m$
is closed with respect to reduction of access patterns.

\begin{lemma}\label{lemma:closure}
  For $\alpha \in L(\nfa_m)$, if $\alpha \rightstar
  \alpha'$ and $\alpha' \not= \bot$, then $\alpha' \in  L(\nfa_m)$.
\end{lemma}
\begin{proof}
   Assume $\alpha \rightk{k} \alpha'$. The Proof is by induction on
  $k$, number of steps in reduction.\\
    \noindent{({\em   BASE})}  case  $k=0$   is  trivial   as  $\alpha
      \rightk{0} \alpha$.\\
    \noindent{({\em   HYPOTHESIS})}  Assume   that  for   $\alpha  \in
      L(\nfa_m)$, if $\alpha  \rightk{k-1} \alpha'$, then $\alpha' \in
      L(\nfa_m)$.\\
    \noindent{({\em  INDUCTION})}   $\alpha  \in  L(\nfa_m)$,  $\alpha
      \rightk{k} \alpha'$. There  exists $\alpha''$ such that: $\alpha
      \rightk{k-1}\;  \alpha''\;  \rightk{1}  \alpha'$.  By  induction
      hypothesis, we have $\alpha'' \in L(\nfa_m)$.

      For $\alpha'' \rightk{1} \alpha'$ to hold we must have $\alpha''
      = \alpha_1\bcar\acar\alpha_2$  and $\alpha' = \alpha_1\alpha_2$,
      or  $\alpha''  =   \alpha_1\bcdr\acdr\alpha_2$  and  $\alpha'  =
      \alpha_1\alpha_2$.   Consider   the   case  when   $\alpha''   =
      \alpha_1\bcar\acar\alpha_2$.   Any  path  in $\nfa_m$  accepting
      $\alpha''$ must have the  following structure ({The states shown
      separately may not necessarily be different}):
      \begin{center}
	\scalebox{1}
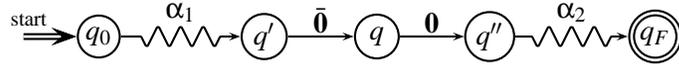
{\begin{pspicture}(0,0)(10,1)
	\psset{unit=1mm}
	\putnode{s0}{origin}{3}{4}{}
	\putnode{s1}{s0}{10}{0}{\pscirclebox[framesep=.5]{$q_0$}}
	\putnode{s2}{s1}{22}{0}{\pscirclebox[framesep=.5]{$q'$}}
	\putnode{s3}{s2}{15}{0}{\pscirclebox[framesep=1.2]{$q$}}
	\putnode{s4}{s3}{15}{0}{\pscirclebox[framesep=.5]{$q''$}}
	\putnode{s5}{s4}{22}{0}{\pscirclebox[doubleline=true,
	    framesep=.3]{$q_F$}}
	\ncline[doubleline=true]{->}{s0}{s1}
	\aput[.5](.1){\footnotesize start}
	\ncline{->}{s2}{s3}
	\Aput[.2]{\bcar}
	\ncline{->}{s3}{s4}
	\Aput[.2]{\acar}
	\nczigzag[coilarm=2,coilwidth=3,linearc=.2]{->}{s1}{s2}
	\Aput[1.7]{$\alpha_1$}
	\nczigzag[coilarm=2,coilwidth=3,linearc=.2]{->}{s4}{s5}
	\Aput[1.7]{$\alpha_2$}
      \end{pspicture}}
      \end{center}
      As $\nfa_m$  is the  fixed point NFA  for the  iteration process
      described    in    the algorithm,   adding    an
      $\epsilon$-edge between  states $q'$  and $q''$ will  not change
      the  language  accepted by  $\nfa_m$.  But,  the access  pattern
      accepted after adding  an $\epsilon$-edge is $\alpha_1\alpha_2 =
      \alpha'$. Thus, $\alpha' \in L(\nfa_m)$. The case when $\alpha''
      = \alpha_1\bcdr\acdr\alpha_2$ is identical.
\end{proof}

\begin{corollary}\label{corr:closure}
  For  $\alpha \in  L(\nfabar)$,  if $\alpha  \rightstar \alpha'$  and
  $\alpha' \not= \bot$, then $\alpha' \in L(\nfa_m)$.
\end{corollary}
\begin{proof}
  $L(\nfabar)    \subseteq    L(\nfa_m)    \Rightarrow   \alpha    \in
  L(\nfa_m)$.     The     proof      follows     from
  Lemma~\ref{lemma:closure}.
\end{proof}

The following  theorem asserts  the equivalence  of  \nfabar\ and
\nfa\ with respect to the equivalence of access patterns, i.e.\ every
access pattern in \nfabar\ has an equivalent canonical access
pattern in \nfa,  and for every canonical access  pattern in \nfa,
there exists an equivalent access pattern in \nfabar.

\begin{theorem}\label{thm:equiv-nfa}
  Let  \nfabar\ be  an NFA with  underlying  alphabet $\lbrace  \acar,
  \acdr,  \bcar,  \bcdr\rbrace$.   Let  NFA  \nfa\  be  the  NFA  with
  underlying  alphabet $\lbrace \acar,  \acdr\rbrace$ returned  by the
  algorithm.  Then,
  \begin{enumerate}
  \item  if $\alpha  \in L(\nfabar)$,  $\beta$ is  a  canonical access
    pattern such that  $\alpha\rightstar\beta$ and $\beta \not= \bot$,
    then $\beta \in L(\nfa)$.
  \item if  $\beta \in  L(\nfa)$ then there  exists an  access pattern
    $\alpha \in L(\nfabar)$ such that $\alpha\rightstar\beta$.
  \end{enumerate}
\end{theorem}

\begin{proof}
  \begin{enumerate}
  \item    From   Corollary~\ref{corr:closure}:    \mbox{$\alpha   \in
    L(\nfabar),  \alpha\rightstar\beta \mbox{ and  } \beta  \not= \bot
    \Rightarrow  \beta \in  L(\nfa_m)$}.  As  $\beta$ is  in canonical
    form,  the path accepting  $\beta$ in  $\nfa_m$ consists  of edges
    labeled $\acar$ and  $\acdr$ only.  The same path  exists in \nfa.
    Thus \nfa\ also accepts $\beta \Rightarrow \beta \in L(\nfa)$.
  \item $L(\nfa) \subseteq L(\nfa_m) \Rightarrow \beta \in L(\nfa_m)$.
    Using   Corollary~\ref{corr:no-new},  there  exists   $\alpha  \in
    L(\nfabar)$ such that $\alpha\rightstar\beta$.
  \end{enumerate}
\end{proof}

\section{An Application of Liveness Analysis}
\label{sec:apps}

The result of liveness analysis can  be used to decide whether a given
access  path $v.\alpha$  can be  nullified  at a  given program  point
$\pi$.  Let the  link corresponding to $v.\alpha$ in  the memory graph
be $l$.   A naive  approach is  to nullify $v.\alpha$  if it  does not
belongs to  the liveness environment at $\pi$.   However, the approach
is  not safe because  of two  reasons: (a)  The link  $l$ may  be used
beyond $\pi$ through  an alias, and may therefore be  live. (b) a link
$l'$ in  the access path  from the root  variable $v$ to $l$  may have
been created  along one execution  path but not along  another.  Since
the  nullification  of  $v.\alpha$   requires  the  link  $l'$  to  be
dereferenced, a run time exception may occur.

To solve the first problem, we  need an alias analysis phase to detect
sharing of links  among access paths.  A link in  the memory graph can
be nullified at $\pi$ if none  of the access paths sharing it are live
at  $\pi$ .   To solve  the second  problem, we  need  an availability
analysis phase.  It detects whether  all links in the access path have
been created along all execution  paths reaching $\pi$. The results of
these  analysis  are used  to  filter  out  those access  paths  whose
nullification may  be unsafe.  We  do not address the  descriptions of
these analyses in this paper.

\section{Related Work}
\label{sec:rel-work}

In this paper, we have described a static analysis for inferring
dead  references in  first  order functional  programs.   We employ  a
context free  grammar based abstraction for  the heap. This  is in the
spirit  of  the  work  by Jones  and  Muchnick~\cite{jones79flow}  for
functional programs.
The  existing literature  related  to improving  memory efficiency  of
programs can be categorized as follows:

{\em Compile time  reuse}.  The method by Barth~\cite{barth77shifting}
detects memory  cells with zero  reference count and  reallocates them
for     further     use     in     the     program.      Jones     and
Le~Metayer~\cite{jones89compile}  describe  a  sharing analysis  based
garbage collection for reusing  of cells.  Their analysis incorporates
liveness  information: A  cell is  collected  even when  it is  shared
provided expressions sharing it do not need it for their evaluation. 

{\em         Explicit        reclamation}.          Shaham        et.\
al.~\cite{shaham05establishing}  use  an  automaton called  {\em  heap
safety automaton\/} to model safety of inserting a free statement at a
given   program    point.    The   analysis   is    based   on   shape
analysis~\cite{sagiv99shape,sagiv02shape}  and is  very  precise.  The
disadvantage of the analysis is  that it is very inefficient and takes
large time even for toy programs.
{\em   Free-Me}~\cite{guyer06free}  combines  a   lightweight  pointer
analysis  with  liveness  information  that detects  when  short-lived
objects die and  insert statements to free such  objects. The analysis
is  simpler and  cheaper as  the  scope is  limited. 
The  analysis  described   by  Inoue  et.\  al.~\cite{inoue88analysis}
detects the scope (function) out  of which a cell becomes unreachable,
and claims the cell using an explicit {\em reclaim} procedure whenever
the execution goes out of that  scope.  Like our method, the result of
their  analysis is also  represented using  CFGs. The  main difference
between their work  and ours is that we detect  and nullify dead links
at any  point of  the program, while  they detect and  collect objects
that are unreachable at function boundaries.
Cherem and Rugina~\cite{cherem06compile}  use a  shape analysis
framework~\cite{hackett05region} to  analyze a  single heap cell  at a
time for  deallocation. However,  multiple iterations of  the analysis
and the  optimization steps are  required, since freeing a  cell might
result in opportunities for more deallocations.

{\em Making dead objects unreachable}.
The  most popular  approach to  make  dead objects  unreachable is  to
identify live variables in the program  to reduce the root set to only
the   live  reference   variables~\cite{agesen98garbage}.   The  major
drawback of this approach is  that all heap objects reachable from the
live root variables are considered live,  even if some of them may not
be used by the program.
{\em                          Escape                         analysis}
\cite{Blanchet:1999:EAO,Blanchet:2003:EAJ,choi99escape}          based
approaches discover  objects escaping a procedure  (an escaping object
being an  object whose lifetimes  outlives the procedure  that created
it). All  non-escaping objects are  allocated on stack,  thus becoming
unreachable whenever the creating procedure exits.
In  {\em Region}  based garbage  collection~\cite{tofte02combining}, a
static  analysis called {\em  region inference\/}~\cite{tofte98region}
is  used to  identify {\em  regions\/} that  are storage  for objects.
Normal memory blocks  can be allocated at any point  in time; they are
always allocated in a particular region and are deallocated at the end
of that  region's lifetime.  Approaches  based on escape  analysis and
region  inference detect  garbage only  at the  boundaries  of certain
predefined areas of the program.
In  our  previous  work~\cite{khedker06heap},  we  have  used  bounded
abstractions of  access paths called  {\em access graphs}  to describe
the liveness of memory links in imperative programs and have used this
information to nullify dead links.

A related  work due to Heine  and Lam~\cite{heine03practical} attempts
to  find potential  memory leaks  in C/C++  programs by  detecting the
earliest point in a program when an object becomes unreachable.

\section{Conclusions}
\label{sec:concl}

In this  paper we  presented a technique  to compute liveness  of heap
data in functional programs. This information could be used to nullify
links in heap memory to improve garbage collection. We have abstracted
the liveness information in the form of a CFG, which is then converted
to NFAs. This conversion implies  some imprecision. We present a novel
way  to simplify  the  NFAs so  they  directly describe  paths in  the
heap.    Unlike     the    method    described     by    Inoue    et.\
al.~\cite{inoue88analysis},  our  simplification  does not  cause  any
imprecision.

In future, we intend to take  this method to its logical conclusion by
addressing  the issue  of  nullification.  This  would  require us  to
perform  alias  analysis  which we  feel  can  be  done in  a  similar
fashion. We also  feel that with minor modification  our method can be
used for  dead code elimination and  intend to extend  our analysis to
higher order languages.

\bibliographystyle{entcs}

\clearpage
\appendix
\section{Computation of Liveness for \append}
\label{app:compute-app}
\newcommand{\lbeleven}{\mbox{\listb.\xf{\append}{2}($\bcdrset\cdot\sigma$)}}
\newcommand{\leleven}{$\lbrace \lbeleven \rbrace$}
\newcommand{\laten}{\mbox{$\lista.\acdrset\cdot\xf{\append}{1}(\bcdrset\cdot\sigma)$}}
\newcommand{\lten}{$\left\lbrace\begin{array}{c}\laten,\\ \lbeleven\end{array}\right\rbrace$}
\newcommand{\laseven}{\mbox{$\lista.(\epsilonset \cup \lbrace\acar\bcar\rbrace\cdot\sigma \cup \acdrset\cdot\xf{\append}{1}(\bcdrset\cdot\sigma))$}}
\newcommand{\lseven}{$\left\lbrace\begin{array}{c}\laseven,\\ \lbeleven\end{array}\right\rbrace$}
\newcommand{\lbfour}{\mbox{$\listb.(\sigma \cup \xf{\append}{2}(\bcdrset\cdot\sigma))$}}
\newcommand{\lfour}{$\left\lbrace\begin{array}{c}\laseven,\\ \lbfour\end{array}\right\rbrace$}

\begin{center}
\rotatebox{90}{\scalebox{.95}{
 \renewcommand{\arraystretch}{1.2}
\begin{tabular}{|c|c|c|c|} \hline
  Program & Live Access Patterns & Liveness Environment  & Liveness
  Environment\\ 
  Point  & for $e$ at $\pi$  & after $e$  &  at $\pi$  \\
  ($\pi$) & ($\sigma$) & (\lv) & (\xe{e}{\sigma}{\lv}) \\\hline
  $\pi_1$    &$\sigma$&$\emptyset$&\lfour \\ \hline
  $\pi_2$    &\epsilonset&\lfour&\lfour\\ \hline
  $\pi_3$    &\epsilonset&\lfour&\lfour\\ \hline
  $\pi_4$    &$\sigma$&\lseven&\lfour\\ \hline
  $\pi_5$    &$\sigma$&$\emptyset$&\lseven\\ \hline
  $\pi_6$    &$\bcarset\cdot\sigma$&\lten&\lseven\\ \hline
  $\pi_7$    &$\epsilonset\cup\lbrace\acar\bcar\rbrace\cdot\sigma$&\lten&
  \lseven\\ \hline
  $\pi_8$    &$\bcdrset\cdot\sigma$&$\emptyset$&\lten\\ \hline
  $\pi_9$    &\xf{\append}{1}($\bcdrset\cdot\sigma$)&\leleven&\lten\\ \hline
  $\pi_{10}$ &$\acdrset\cdot\xf{\append}{2}(\bcdrset\cdot\sigma)$&\leleven&
  \lten \\ \hline
  $\pi_{11}$
  &\xf{\append}{2}($\bcdrset\cdot\sigma$)&$\emptyset$&\leleven\\
  \hline 
\end{tabular}}}
\end{center}

\end{document}